\documentclass{llncs/llncs} 
\usepackage[UKenglish]{babel}
\usepackage[utf8]{inputenc}
\usepackage{amsmath}
\usepackage{amssymb}
\usepackage{graphicx}
\usepackage{subfigure}
\usepackage{cite}
\usepackage{amsfonts}   
\usepackage{amssymb}
\usepackage{amsmath}
\usepackage{verbatim}
\usepackage{wrapfig}
\usepackage{pdfpages}

\begin{document}


\title{Algorithms for Timed Consistency Models}
\author{Moritz Schattka}
\institute{Karlsruhe Institute of Technology}
\maketitle

\begin{abstract}
One of the major challenges in distributed systems is establishing consistency among replicated data in a timely fashion. While the consistent \textit{ordering} of events has been extensively researched, the \textit{time span} to reach a consistent state is mostly considered an effect of the chosen consistency model, rather than being considered a parameter itself.
This paper argues that it is possible to give guarantees on the timely consistency of an operation. Subsequent to an update the cloud and all connected clients will either be consistent with the update within the defined upper bound of time $\Delta$ or the update will be returned. This paper suggests the respective algorithms and protocols capable of producing such comprehensive \textit{Timed Consistency}, as conceptually proposed by Torres-Rojas et al.
The solution offers business customers an increasing level of predictability and adjustability. The temporal certainty concerning the execution makes the cloud a more attractive tool for time-critical or mission-critical applications fearing the poor availability of Strong Consistency in cloud environments.
 
\end{abstract}


\thispagestyle{plain}

\section{Introduction}

Cloud computing is continuously transforming the IT landscape for businesses. Driven by the multitude of opportunities to drastically improve efficiency, performance, scalability and reliability, a rapidly increasing number of today's businesses makes use of the cloud\cite{Forrester1}. The cloud can offer great advantages both economically and performance-wise; whether for web-services, storage or even outsourcing of enterprise software and entire infrastructures. The cloud is becoming increasingly popular among companies. While it can be a key value driver for business success, data consistency is one of today's major concerns for developers building cloud-based applications. Data entering the cloud is replicated among different server sites to improve reliability and performance\cite{TB}. These sites can be different servers in a data center or even servers in data centers spread around the globe. While the improvements in performance and reliability offer obvious advantages, they come with the headache of keeping all replicas up-to-date and reachable in a timely fashion. Archiving data consistency in a large-scale system operating on a global scale leads to many trade-offs that need to be considered. 

In order to make efficient use of the cloud, traditional paradigms had to undergo drastic changes. The most desirable system would be one that is "always available for reads and writes, and able to continue operating during network partitions"\cite{COPS}. Accordingly updates would be available to all observers always and immediately. This model of consistency is known as Strong Consistency\cite{Vogels}. In the late 1970s it was becoming apparent that this paradigm needed to be revisited\cite{NoDD}. It was becoming more obvious that Strong Consistency properties were hard to realize in increasingly distributed systems. As systems become larger and more distributed, in both a geographical and technical sense, the number of machines involved in operations increases and with it the necessity for communication over potentially unreliable networks. If the Strong Consistency is suppose to hold, a high number of machines is required to successfully interact on any operation a user performs to keep all replicas updated. Every server storing a copy will have to successfully write the data and confirm the operations. This is not only extremely time consuming with growing network size.
A single break in the communication channels or failure on one of the involved machines would cause Strong Consistency to be broken. The system would have to fail the operation and refuse any further requests until the inconsistency was fixed. Large amounts of data can become unavailable due to this lack of failure tolerance. "Small and large components fail continuously"\cite{ADynamo} and the larger a distributed system grows, the higher the probability of a failure gets. Strong Consistency prevents effective scaling\cite{Ahamad}. 

It is not only hard to ensure Strong Consistency in terms of networks failing and the server infrastructure being globally distributed. Making updates of the data immediately available on all replicas has many implications performance-wise and cost-wise as well. A user who is, through the cloud, connected to one server sight might produce a data overhead by a single write operation that is many dimensions larger than the operation itself. Before he can perform any further write or read operation all sites that store a copy of that particular piece of data need to be updated and confirm the operation. This implies great amounts of data being written all across the cloud infrastructure and a great quota of network traffic being utilized. 

The approach of rather failing the system than breaking consistency was fundamentally changed with large internet systems coming up in the 1990s. The classic ACID properties (Atomicity, Consistency, Isolation, Durability) developed by Jim Gray in the 1970's\cite{ACID1} to describe reliable transactions in databases just did not produce sufficient  availability in distributed data stores\cite{ADynamo}\cite{antiacid}. "Consistency [...] can not longer be guaranteed in a distributed system, where concurrent operations are occurring"\cite{CAP}.

The idea of replacing consistency by availability as the most important characteristic of a system was getting increasingly popular. In his keynote to the PODC (Principles of Distributed Computing) conference in 2000 Eric Brewer addresses this development by formulating the trade-off a developer is facing in his famous CAP-Theorem\cite{CAP}. As stated by Brewer and later formally proven by Gilbert and Lynch\cite{CAP-Proof}, of the three desired properties of a shared data-system - consistency, availability, tolerance to network partition - only two can be simultaneously guaranteed. Weakening the consistency constraint offered a chance to develop highly available systems while reducing latency and facilitated the scaleability of the systems\cite{COPS}. 

The sacrifice of Strong Consistency triggered the development of different consistency models. Those implemented the idea to trade off immediate and complete consistency against guaranteeing higher valued properties. Most of the models can be summed up under the model of \textit{Eventual Consistency}. It guarantees that "if no updates take place for a long time, all replicas will gradually become consistent"\cite{TB}. The consistency model applied varies depending on the intended use. While for example banking applications obviously require a very strong consistency model to avoid the same unit of currency being withdrawn multiple times; a weaker model will be totally sufficient  for a news feed.

Although there is a wide range of variations to the model of Eventual Consistency (see \cite{Vogels}) they might not address one vital interest. While the user might not necessarily require the data written to the system to be instantly consistent, he might very well require to know \textit{when} the data is consistent. Put differently, a user requiring temporal guarantees might find it much more natural to have the consistency guarantees expressed temporarily\cite{Delta}. To provide this guarantee to the user \textit{Francisco Torres-Rojas et al.} propose an approach called a Timed Consistency Model\cite{TR}. It "defines a maximum acceptable threshold of time after which the effect of a write operation must be observed by all the sites of a distributed system"\cite{TR}. 

Timed Consistency has the potential to address the user's need to know about the staleness of the data he reads. In addition the maximum acceptable threshold of inconsistency, the length of the inconsistency window, can be defined according to the needs of the customer. Depending on his requirements he can set the upper bound for the inconsistency window to a value that best represents his preferred trade-off between consistency, availability and cost. To give effective guarantees to business customers, these measurable parameters can be agreed upon in Service Level Agreements (SLA).

This paper provides an overview on the different models of Timed Consistency, the related theoretical paradigms and suggests a possible implementation for Timed Consistency. The implementation is described with respective algorithms and protocols. Finally the feasibility of Timed Consistency will be evaluated from a technical and an economic point of view and it will be discussed whether Timed Consistency can effectively address business customer's needs.

\section{Review of Consistency Models}
There are two reasons for tolerating data inconsistency in a distributed system:
To enhance availability and to improve performance\cite{TB}. Weakening consistency allows handling network partition cases where strong consistency would "render parts of the system unavailable eventhough the nodes are up and running"\cite{Vogels}. Also, under highly concurrent conditions a degree of tolerance improves read and write performance. \cite{Vogels}

"Consistency Models aim at providing a systemwide consistent view on a data store"\cite{TB}.
They weaken consistency and still allow a user to read consistent data if certain requirements are met.
Consistency Models can in that sense be interpreted as contracts between the user and the system. Given that a user obeys certain rules the system will behave consistently. Having knowledge of the consistency model applied is important to make the system predictable for the developer using the cloud. There is a variety of models that differ in the way they weaken consistency and that conclusively model the trade-off between consistency and other properties differently. Depending on the desired application different models are preferable. Furthermore, the implementation complexity drastically varies depending on the applied model.

\subsection{Introduction to Consistency Models}

"Ordering and time are two different aspects of consistency of shared objects in a distributed system. One avoids conflicts between operations, the other addresses how quickly the effects of an operation are perceived by the rest of the system"\cite{TR}. There is two points of view to evaluate consistency: \textit{Client-centric} and \textit{data-Centric}\cite{TB}.
Client-centric consistency models define how users observe updates of the data and whether or not the read data is stale, while not taking into account the internal state of a storage system. The data-centric side on the other hand is concerned with the internal state of a storage system, observing whether the stored replicas are identical. \cite{David1}

As already demonstrated, a consistency model is either \textit{strong}, meaning after an update all read requests return the same result, or \textit{weak}, meaning the system guarantees consistency only under certain conditions.
Eventual Consistency is a specific form of weak consistency guaranteeing that all updates will \textit{gradually} become consistent if no updates take place for a long time\cite{Vogels}\cite{TB}. The Eventual Consistency Model has a number of variations\cite{Vogels}: 

\subsubsection{Sequential Consistency (SC).}
Every machine sees write operations on a single piece of data in the same sequential order, although the operations might not necessarily have occurred in this specific temporal order\cite{Lamport}. As it is essentially impossible to have all users clocks sufficiently synchronized to produce a valid temporal order, SC will at least avoid any further problems by producing a consistent order. \cite{Lamport}

\subsubsection{Causal Consistency(CC).}

CC guarantees that every \textit{causally related} operation will be consistent\cite{Ahamad}. Only those events that have the potential of being causally related need to be in the same sequential order on all machines. For those events that are concurrent it is tolerated to have a different sequential ordering on the different machines\cite{TB}. This effectively avoids conflicts while being comparably weaker than SC and thus improves performance and availability.

\subsection{Timed Consistency}

Timed Consistency Models were introduced by \textit{Torres-Rojas et al.} in their 1999 paper \textit{"Timed Consistency for Shared Distributed Objects"} \cite{TR}. Timed Consistency supplements existing consistency criteria by one important aspect: Any operation executed at time \textit{t} must be visible to every user at the latest by time \textit{t+$\Delta$}.

The $\Delta$ defines the "maximum acceptable threshold"\cite{TR} for the inconsistency window. After the inconsistency window has ended any user must be able to observe the update. If $\Delta$ is 0 the requirements for Strong Consistency are fulfilled. \textit{Torres-Rojas et al.} define two variations of Timed Consistency that extend the popular models of Causal Consistency and Sequential Consistency by the requirement of reading on time. "Neither Sequential Consistency, nor Causal Consistency consider the particular time an operation is executed. Their goal is to establish a valid order among all the operations"\cite{TR}. 
Once extended by the need of reading within $\Delta$ they are called \textit{Timed Causal Consistency (TCC)} and \textit{Timed Sequential Consistency (TSC)}.\cite{TR}.  \textit{Torres-Rojas et al.} propose an approach using the so-called \textit{Lifetime Based Consistency Protocol} to make SC and CC "timed".

\subsubsection{Lifetime Based Consistency Protocol(LBCP).}
The \textit{Lifetime Based Consistency Protocol} is a model to archive Timed Consistency in a distributed system. It is widely found in read driven applications with comparably low volumes. The most common application is the \textit{Website Cache} of a browser that locally stores copies of frequently used websites to avoid downloading the same content from a web server multiple times and thus improves latencies and reduces network loads.

The LBCP requires that a user reading data from a remote source (e.g. cloud) needs to store a copy of this data locally (e.g. hard drive)\cite{Lifetime}. Any object a user caches gets a time stamp with a \textit{start time} and an \textit{ending time}. The \textit{start time} corresponds to the time the write to the local cache is successfully completed. The \textit{ending time} defines the latest time a value stored in the cache is still valid. These two values correspond to beginning and end of the \textit{lifetime} of an object. Now, if the user wants to read a certain object he first accesses his local cache to check if he has a local copy in storage and to verify whether his local copy of the object is still valid; this means that that the current time is \textit{before} the object's ending time.

Two different objects are called \textit{mutually consistent} when there is a time \textit{t} at which both objects are valid. The cache can be called consistent if all objects are mutually consistent at a time \textit{t}, meaning the latest starting time of all objects is smaller than the earliest ending time of all the objects\cite{TR}.

\subsubsection{Timed Sequential Consistency (TSC).}
\textit{Timed Sequential Consistency} is based on \textit{Sequential Consistency} and supplements it by the aspect of time. Any operation needs to meet the requirements of SC after time $\Delta$. As it does not require immediate consistency to keep the system available but instead leaves a time window of $\Delta$ for the system to become consistent it is considered weaker than SC\cite{TR}.

\textit{Torres-Rojas et al.} propose using the caching method of LBCP to induce Timed Sequential Consistency in their 1999 paper \textit{Timed Consistency for Shared Distributed Objects}\cite{TR}. Objects in the cache that are older than $\Delta$ are automatically invalidated locally. If the lifetime of an object corresponds to the time span defined as $\Delta$ it is ensured that after the time of $\Delta$, which is potentially an inconsistency window if a write has occurred in the mean time, all objects have returned to a consistent state. A write operation at time \textit{t} will be visible, meaning updated in the cache, to all users latest by \textit{t+$\Delta$}. By then all users have refreshed their cached object.
While this method is efficient for regularly or even periodically changing objects that are accessed often, it can create a rather big overhead for objects that are either hardly ever updated or not accessed by the user very often\cite{TR}. Therefore \textit{Torres-Rojas et al.} proposes not to \textit{invalidate} the objects but rather to mark those objects that have passed their lifetime as \textit{old}. Access to these objects by the user will then create an \textit{if-modified-since} request to the distributed system. If the object has been updated in the meanwhile it will be updated in the user's cache, otherwise the object in the local cache will again be validated for a lifetime of $\Delta$. This can drastically reduce the need to transfer large objects.

For further analysis the cache will be considered part of the data-side as its function is to provide the data for read operations that otherwise would need to be requested from the distributed system. 

\subsubsection{Clocks.}
Clocks are essential for a distributed system and especially for TSC, where both order and time are highly relevant. While order is important to get the sequence of events right and avoid conflicts, time needs to be measured to guarantee the maximum threshold $\Delta$ for an update to flow through the system of clients and cloud.

For two processes which are causally \textit{unrelated} and thus do not interact, it is irrelevant whether or not they are executed in a correct order compared to real time. As the result is going to be the same, independent of their order, the difference could not even be observed\cite{Lamport2}. Those processes are called \textit{concurrent}\cite{TB}. The situation is very different for causally related operations. In order to correctly execute them, it is essential to know their correct sequential order. If the order can not be determined correctly it endangers the correct causal execution and thus might cause serious conflicts. The system will need some kind of clock to determine the order of objects to correctly process them. 
One option is using so called physical clocks, measuring real time. 
Physical clocks are based on quartz crystals in form of integrated circuits. These produce a steady pace which then is transformed into a time value. The great disadvantage of physical clocks is that they are requiring every server and, depending on the protocols, even all clients, to have almost perfectly synchronous clocks. Unfortunately physical clocks have the tendency of drifting apart due to minimal differences in there physical properties and therefore need continuous resynchronization\cite{lamport3}. On a spatially distributed network where the runtimes of messages may vary significantly, this is not possible without leaving a big margin of tolerance.\cite{bto} This tolerance has to be taken care of in the respective algorithms, which can drastically affect performance.

As \textit{Lamport} pointed out in his 1978's paper\cite{Lamport2} it is not necessary for processes to agree on the time but rather to agree on the order in which events occur\cite{TB} and as we have already seen ordering and time are two different aspects of consistency\cite{TR}.
\textit{Lamport} addresses this problem by suggesting a logical clock called the Lamport Timestamp\cite{Lamport2}. Logical Clocks do not have the purpose to measure physical time but rather to give consecutive events a monotonously increasing value as a timestamp. The Lamport Timestamp is a certain logical clock that makes it possible to give events a partial causal ordering. Let \textit{A, B, C}  (See Figure 1) be three different processes with their respective events (e.g. \textit{A1, B4, C2}). Every process has an own clock starting from 0. Now, whenever a process accesses an object it writes the present value of its clock into the object's timestamp \textit{n}. Another process reading from this shared object will read the timestamp and compare it to its own process's timestamp. If the objects timestamp carries a \textit{higher value} than the process's clock it will set it's own clock to \textit{n+1}, with \textit{n} being the object's timestamp. If \textit{n} is \textit{lower} than the process's clock it will synchronize $n$ to its local clock and add the value of one to both to acknowledge the process.

These rules ensure that if two events \textit{are} causally dependent they can be put in the right order just by sorting them by their timestamp's value (\textit{Weak Clock Consistency Condition}). The reverse is not the case. By ordering event's timestamps, conclusions on the causality can \textit{not} be made.
Lamport Clocks are unable to detect causalities, so without further knowledge on the causality of events, the Lamport Clock will not be able to generate a valid order among a random set of events.

\begin{figure}
\includegraphics[height=60mm]{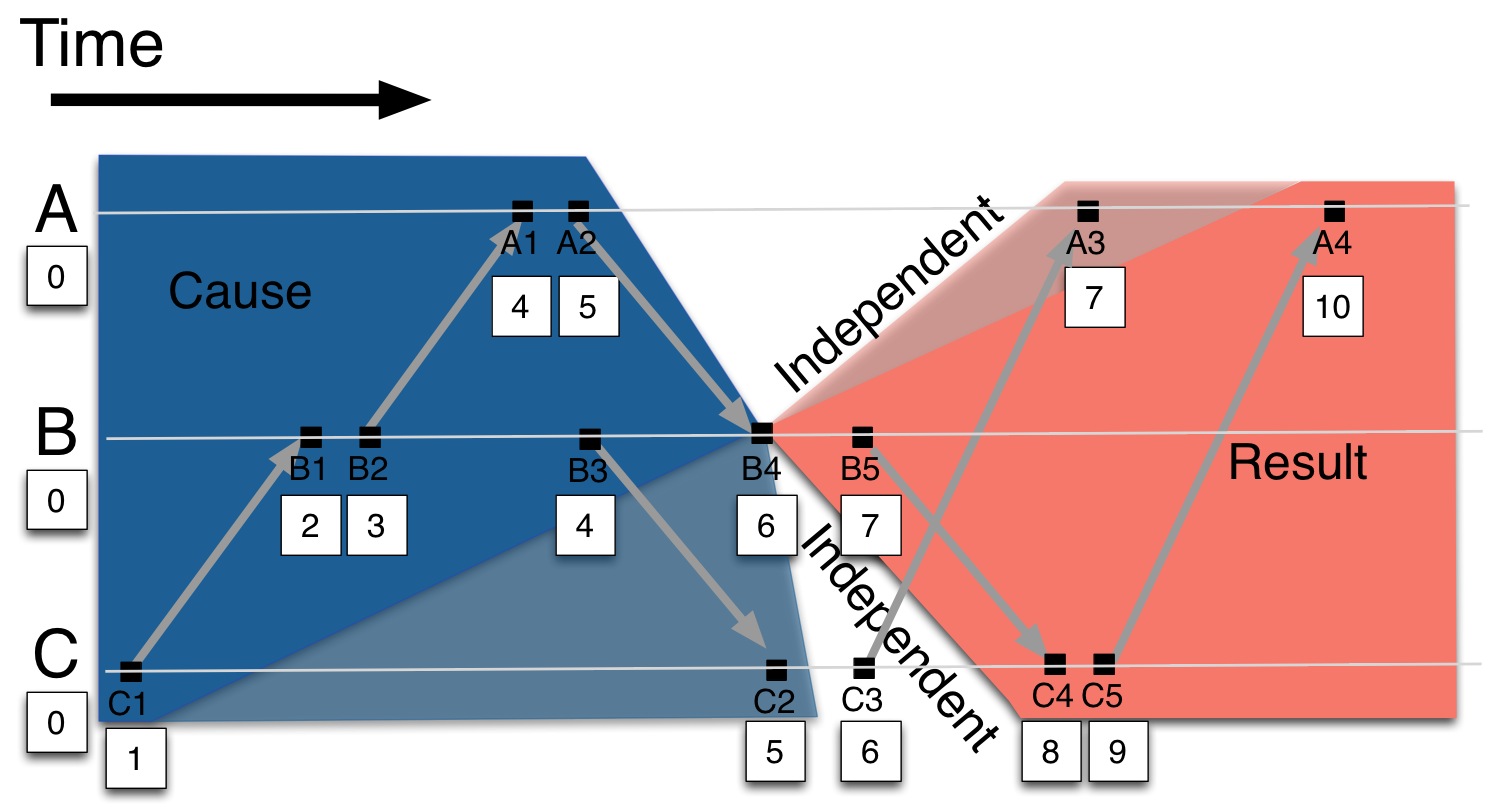}
\caption{Lamport Clock with process A, B, C. "Events in the blue region are the causes leading to event B4, whereas those in the red region are the effects of event B4"\cite{imgvector}}
\end{figure}

 To solve this issue \textit{Vector Clocks} were developed based on \textit{Lamport's} concept (see \cite{fidge}\cite{mattern}). Contrary to his model Vector Clocks do not only have a timestamp with \textit{one} dimension but a timestamp vector with \textit{N} dimension, \textit{N} being  the number of processes in the distributed system. Every time a process experiences an event it increases its own clock by one. It writes this value into the respective column of the object's timestamp vector. As soon as another process accesses the object it also writes its clock's value into the respective column of the timestamp vector. By reading the vector a process can determine how many events have occurred in any other process previously and \textit{may} have causal influence on the objects. Any object with a timestamp with all the components of the vector being \textit{smaller or equal} to another object's timestamp are potentially causal (see Figure 2). This way vector clocks offer a tool to detect concurrency and dependency.

Formally the causal relationship of an event \textit{x} happening before an event \textit{y} can be written as: $x\to y$ \cite{Lamport}. \textit{C(x)} denotes the timestamp of an event \textit{x}.
For a logical clock allowing to put \textit{causally related} and \textit{concurrent} events in a 
correct sequential order by comparing their timestamps the  \textit{Strong Clock Consistency Condition} needs to be met:
 $$\ C(x) \leq \ C(y) \Longleftrightarrow x\to y$$
 
While Vector Clocks meet the Strong Clock Consistency Condition, Lamport Clocks only meet the \textit{"Weak Clock Consistency Condition"}:

 $$\ C(x) \leq \ C(y) \Rightarrow x\to y$$

Two Vector Timestamps can be compared by the following formula:

\[
\displaystyle C(x)<C(y) \iff  \forall z[ C_z(x)\leq C_z(y) ]   \land \exists z' [ C_{z'}(x)< C_{z'}(y) ] 
\]

The value \textit{C(x)} of a Vector Clock \textit{x} is smaller then the value \textit{C(y)} of a Vector Clock \textit{y} if, and only if, the entries in all rows $C_z(x)$ of vector \textit{x} are smaller or equal to the entries in the corresponding rows $C_z(y)$ of vector \textit{y} and an entry $C_{z'}(x)$ exists in C(x) that is smaller than the corresponding entry in C(y).

\begin{figure}[h]
\includegraphics[height=60mm]{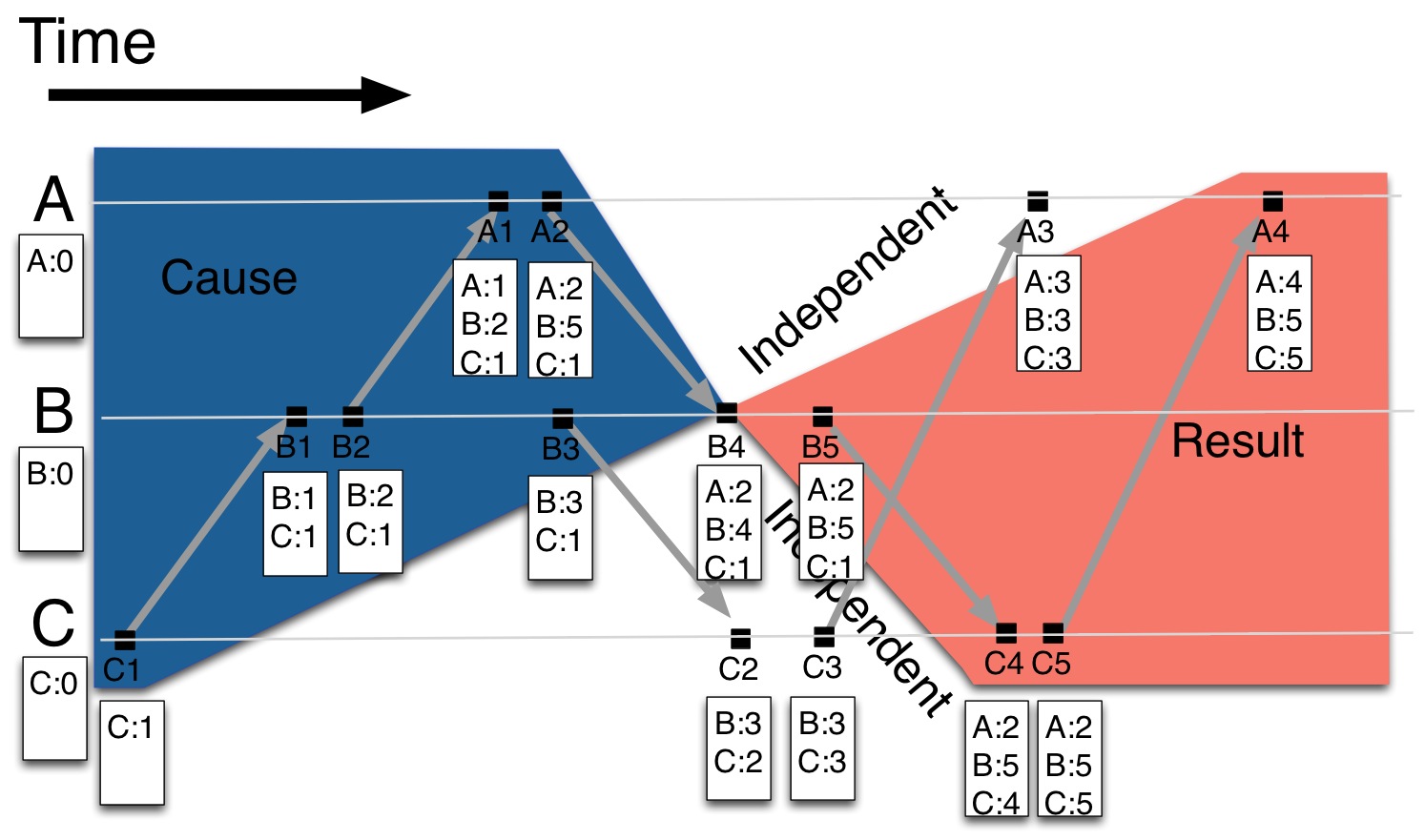}
\caption{Vector Clock with process A, B, C. "Events in the blue region are the causes leading to event B4, whereas those in the red region are the effects of event B4"\cite{imgvector}}
\end{figure}

\subsubsection{Timed Causal Consistency (TCC).}
Analog to TSC, \textit{Torres-Rojas et al.} base Timed Causal Consistency on the Lifetime Based Consistency Protocol\cite{CTaWCM}. Opposite to TSC, Timed Causal Consistency will waive the measurement of $\Delta$ in realtime and rather use an approach based on logical time. $\Delta$ is not longer being measured in time units but as a real number. Instead of computing $\Delta$ as the difference of the real time and the timestamp of an object, the Vector Clock's timestamps will be compared. The $\Delta$ is now defined as the difference in events a timestamp is "aware" of\cite{TR}. \textit{Timed} now means that an object will be consistent on a machine at the latest after a certain number $\Delta$ of events has occurred, instead of after a certain amount of real time has passed.
The awareness horizon $\xi_i$ of a timestamp \textit{i} is defined as the sum of the entries of its vector.

\[
\displaystyle \xi_i= \sum_{i=0}^{N-1} t[i]
\]

In order to determine the difference $\Delta_i$ of the awareness horizon $\xi_s$ of timestamp $C(s)$ to the horizon $\xi_i$ of the timestamp $C(i)$ one only needs to compute the difference $\Delta_i$ as followed:

\[
\displaystyle \Delta_i=\xi_s-\xi_i
\]
An object is now valid if the maximum allowed inconsistency window $\Delta$ is bigger than the actual difference in timestamps $\Delta_i$.

\[
\displaystyle\xi_s-\xi_i = \Delta_i\leq\Delta
\]
For instance if $s=(10,22,7,0,3)$ and $i=(2,7,5,0,1)$, then $\xi_s=42$ and $\xi_i=15$. 

\[
\displaystyle\xi_s-\xi_i =42-15=27=\Delta_i
\]
For any value $\Delta<27$ the object \textit{i} would be invalid.

\section{Implementation of Timed Sequential Consistency}

The LBCP-based approach that \textit{Torres-Rojas et al.} propose in their paper has the goal of implementing TSC. They let the client's cache update stored values in intervals of $\Delta$, expecting this technique to implicate TSC. They argue that if a value in the cache is updated at the latest every $\Delta$, TSC can be assumed as a result. 

As seen before, client-centric consistency describes the way a user experiences consistency. If TSC is implemented client-centric this implies that every update a client performs needs to be consistent at any other client at the \textit{latest} after time $\Delta$. In the LBCP-based approach of \textit{Torres-Rojas et al.} the cache ensures that the client does not read stale data by refreshing content that is older than $\Delta$. This implicates one very basic assumption: All the data the client's cache loads from the distributed system imperatively \textit{needs} to be consistent across the distributed system at the time of the read. The distributed system would therefore be required to transfer updates with latency of zero from a user to all the servers in the distributed system to guarantee the update the client requests is not stale. As this is not realistic the result of such an implementation would be unsatisfying. It is possible that the cache sends a read request to the distributed system to update a stored value and that the distributed system returns a stale value because the certain machine the cache is connected to is not yet consistent internally or with the client triggering the update. The key concern for proposing an implementation based on the paper of \textit{Torres-Rojas et al.} is, that it simply does not consider the implications of data-centric consistency.
\begin{wrapfigure}{r}{0.5\textwidth}
  \vspace{-30pt}
  \begin{center}
    \includegraphics[height=70mm]{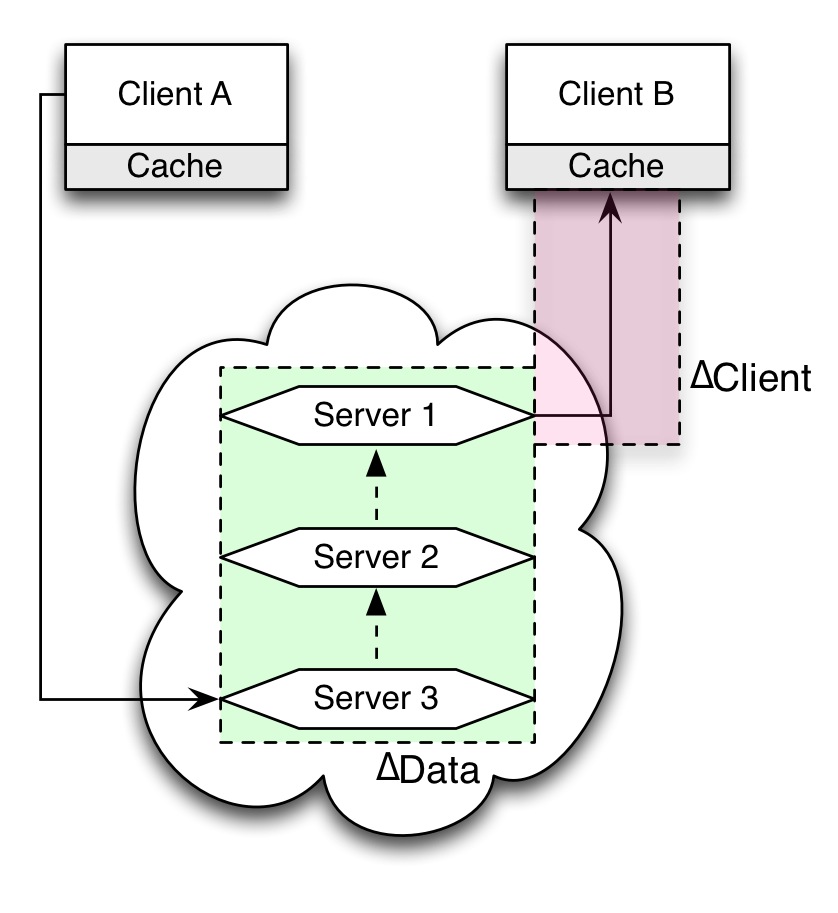}
  \end{center}
  \vspace{-30pt}
\caption{An update flowing through the system with corresponding time intervals $\Delta_{Data}$ and $\Delta_{Client}$.}

  \vspace{-20pt}
\end{wrapfigure}

To resolve this issue the solution proposed in this paper goes further with its implementation of the \textit{timed} aspect. Instead of defining $\Delta$ as the time required for an update to move from the source (e.g. distributed system) to the client's cache, as Torres-Rojas does, $\Delta$ will have to be defined as the maximum time necessary for an update to travel from one client writing the update to another client receiving it. The very difference lies in the data-centric consistency. Instead of requiring the simplifying and unrealistic assumption of a strong and immediate consistency in the distributed system, TSC will also be applied to data-centric consistency. The result will be one $\Delta_{Data}$, that describes the time an update needs to flow through the distributed system once it is successfully written by one client, and a $\Delta_{Client}$ that expresses the maximum time required until an update of the distributed system is loaded onto the client's cache. The maximum total time an update can require to travel from one user to another is expressed by $\Delta_{Update}$:
$$\Delta _{Update}=\Delta_{Data}+\Delta_{Client}$$

Using both $\Delta_{Data}$ and $\Delta_{Client}$ offers the advantages of the LBCP, as Torres-Rojas recommends, with the benefits of Weak Consistency and temporal guarantees. As in LBCP, the lower refresh-rate decreases network load and unnecessary request to be handled by the distributed system, while giving a guarantee on how often updates can be received. Applying data-centric TSC gives guarantees on how long it takes for a performed update to reach all machines in the distributed system while being a rather weak consistency that can be adapted based on the respective latency requirements.

\begin{wrapfigure}{r}{0.5\textwidth}
  \vspace{-35pt}
  \begin{center}
 \includegraphics[height=60mm]{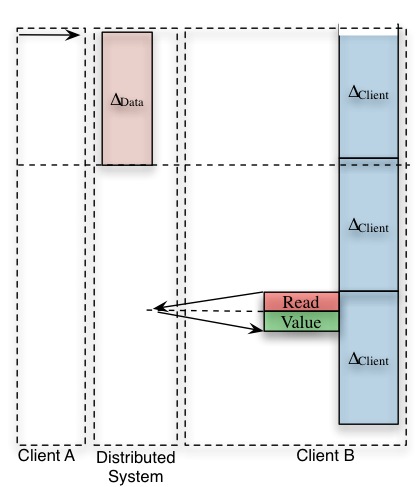}
  \end{center}
  \vspace{-20pt}
\caption{Timeline depicting the maximum dwell time for an update.}
  \vspace{-10pt}
\end{wrapfigure}

As depicted in the example in \textit{Figure 3} the time an update requires to travel from the client performing the write (Client A) to his respective server is considered to be zero. As the update is \textit{pushed} to the server there is no other delay than latency to transfer it. To avoid that this latency affects the guarantees that can be given, it will be defined that the point in time of a particular \textit{write} is the moment the information has finished being processed by the server. The client writing cannot exactly determine when this point in time will be but the transfer protocol makes the exact knowledge unnecessary. The Transmission Control Protocol (TCP) which is the most commonly used protocol to exchange data in web applications confirms the transfer of data by sending back a confirmation message\cite{tcp}. This message can only be send after the server has received the write request. For the client this means that once it received the TCP confirmation message the $\Delta_{Data}$ has already started and that this data package will be reaching any other user at the \textit{latest} by $\Delta _{Update}=\Delta_{Data}+\Delta_{Client}$.

The time $\Delta_{Client}$ does not capture the network latency for the transfer of an update from the distributed system to the client, but the interval in which the client updates (see LBCP). In order to correctly calculate $\Delta_{Update}$ it is necessary to take into account the latency of the cache \textit{sending} a read request and receiving the requested \textit{value} as well as the refresh-rate of the cache performing \textit{if-modified-since} requests.
The value for $\Delta_{Update}$ in a worst case, as depicted in Figure 4, would be:
$$\Delta_{Update}=\Delta_{Data}+\Delta_{Client}+L_{Read}+L_{Value}$$

To be able to make guarantees on a client-centric TSC it will be necessary to define an upper bound for both $L_{Read}$ and $L_{Value}$. This would be possible by letting the cache use a timer for the read request. If the client's cache does not receive a value from the distributed system after a arbitrary time of $\Delta_{Network}$ it can simply invalidate itself temporarily. This would make the whole chain, from the user performing the update to any other user receiving it, predictable with an upper bound of:
$$\Delta_{Update}=\Delta_{Data}+\Delta_{Client}+\Delta_{Network}$$

As a conclusion it is possible to guarantee client-centric TSC by implementing a solution based on the upper bounds $\Delta_{Data}$, $\Delta_{Client}$, $\Delta_{Network}$. \\$\Delta_{Client}$ is already defined as an upper bound by \textit{Torres-Rojas et al.} and $\Delta_{Network}$ can easily be defined using a timer as just shown. What will now be required, is to successfully define a $\Delta_{Data}$ that can be met by the distributed system and successfully implements \textit{data-centric} TSC. Developing the structure for a distributed system that can met TSC with a maximum time of $\Delta_{Data}$ (later referred to as $\Delta$) will be the focus for the following implementation

\subsection{Requirements}

As shown, client-centric \textit{Timed Sequential Consistency} can only be generated if it is also possible to develop a distributed system that establishes \textit{data-centric} TSC. Such a system is required to successfully become sequentially consistent within the time window of $\Delta_{Data}$(later referred to as $\Delta$) and handle all kinds of possible exceptions in such distributed infrastructure.

Distributed Systems consist of a number of servers. The distance of these servers can vary from different racks in a data center to different continents. While this improves availability and latency for clients worldwide it implies a number of challenges for the algorithms trying to handle the consistency issues. In such a big network many factors cannot be taken for granted and many variables can hardly be foreseen. The list below is partly based on \textit{"Fallacies of Distributed Computing"}\cite{fallacies} developed by \textit{Bill Joy} and \textit{Tom Lyon}, and advanced by \textit{Peter Deutsch} and \textit{James Gosling}. It lists the most import concerns with regard to designing an appropriate solution:

\begin{enumerate}
\item{\textbf{Network Latencies:} Depending on the geographical location of the servers and the network infrastructure connecting them, connection delays can drastically vary. }
\item{\textbf{Network Reliability:} The interconnection of remote data centers is established over the internet. In a network that is not proprietary, like the internet, and where control over the handling of data is very limited, no guarantees can be given concerning the success of transfers. }
\item{\textbf{Bandwidth limitations:} As the use of the internet infrastructure is open the loads on the network can vary. As a result there is no guarantees of the bandwidth that can be achieved in such connections. }
\item{\textbf{Heterogeneous Infrastructure:} In a distributed system that is not built from scratch it is most unlikely that all servers are exactly equal in type and performance. Network components will also drastically vary in the internet. }  
\item{\textbf{System Loads:} It can not be taken for granted that every system carries an equal workload. Depending on the geographical location and daytime, for example, some nodes may be more frequently accessed.}
\item{\textbf{Changing Topology:} The composition of the infrastructure continuously changes. Whether for scaling reasons, maintenance, replacements or upgrades.}
\item{\textbf{Varying Clock Speed:} Every server is equipped with an internal clock that is generally based on a quartz crystal. The regular oscillation of this crystal is measured. Even though the frequency of the oscillation is stable, the exact frequency might slightly vary between different clocks. For this reason clocks continuously drift apart.}
\item{\textbf{Asynchronous Clocks:} It is impossible to perfectly synchronize clocks that are remote\cite{Lamport2}. Thus sequential order cannot be kept just by putting the events into order by their local times.}
\item{\textbf{Server Failure:} Servers will frequently be subject to failures and thus become unreachable till repairs are completed.}
\item{\textbf{Data Integrity:} Data stored on hard drives can change involuntarily with time and either generate differing or invalid values.    }
\end{enumerate}

\subsection{Overview}

The implementation of TSC which this paper will suggest relies on an atomic virtual structure consisting of five nodes. A node is a virtual storage unit that can store data and send messages to other nodes. It is possible to have multiple nodes on one physical machine in a data center. The five nodes are labeled A-E and are logically arrayed in a circle and virtually interconnected by five edges (Figure 5).

For the suggested solution the number of nodes (\textit{N}) will always be five, although it is possible to use any number $N=2n+1, n \in \mathbb{N}$ of nodes with the algorithm. The number of five nodes has multiple advantages: First of all the algorithm is quorum-based (See \cite{quorum}) meaning every operation needs to be agreed upon by a majority of nodes. To avoid any standoff situations the number of nodes is required to be uneven. The only smaller uneven number than five that would still constitute a distributed system would be three. A single failing node in a circle of three nodes would leave the system relying on two nodes that will continuously have to agree on any read or write because a failing node is always considered as not-agreeing in a quorum decision. This is highly unlikely in a geographically distributed system which is required to have short latencies. As it will later be shown the algorithm is designed to enforce very strict time limitations for network and system processing to guarantee consistency after a very short time window $\Delta$. Small latency variations, as they are very likely to be occurring in a distributed system, would either make the unanimous decisions of the two remaining nodes highly unlikely or it would force the client to tolerate a much higher $\Delta$. Poor availability or speed will be the result.

A number of nodes $N>5$ is possible and would need to be considered depending on the exact application. High numbers of $N$ are usually found in very read-centric applications\cite{Vogels}. Another reason for having five nodes is the distance of any pair of  two nodes. Two nodes are never more remote than two edges in a circle of five. A higher $N$ makes the healing of inconsistencies either slower or increase the network load considerably, as it will be demonstrated later.
\begin{wrapfigure}{r}{0.5\textwidth}
  \vspace{-30pt}
  \begin{center}
\includegraphics[width=65mm]{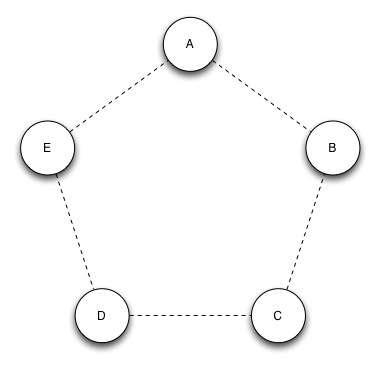}
  \end{center}
  \vspace{-20pt}
\caption{Circle of five nodes A-E}
  \vspace{-10pt}
\end{wrapfigure}

The configuration of a distributed system can be described in terms of replication by the tupel \textit{(N, R, W)}\cite{Vogels}. \textit{N} is, as already explained above, the number of replicas of a piece of data. Every single of this replica lies on a different node. \textit{R} defines the number of nodes required to reply to a \textit{read} request in order for it to be successful. \textit{W} is the number of nodes that needs to confirm the successful completion of a \textit{write} before it is committed to in the distributed system and the confirmation is sent to the client. 
The implementation suggested in this paper uses a configuration of \textit{(N=5, R=3, W=3)} or in the more general form \textit{(2n+1, n+1, n+1)}.

\subsection{Protocol}
\begin{wrapfigure}{r}{0.5\textwidth}
  \vspace{-30pt}
  \begin{center}
\includegraphics[width=60mm]{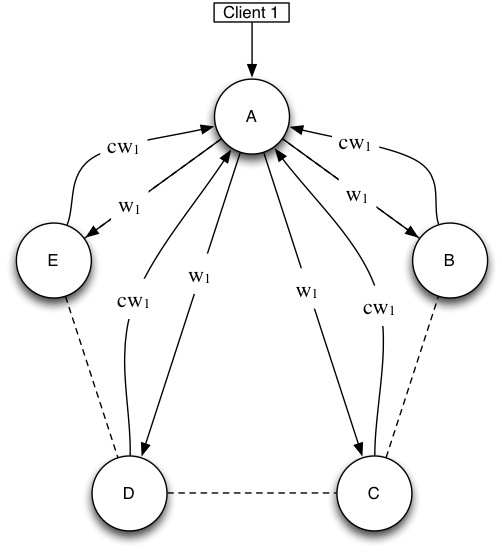}
  \end{center}
  \vspace{-20pt}
\caption{\textit{Client 1} performing a \textit{write} \textit{$w_1$} and receiving confirmations $cw_1$}
  \vspace{-20pt}
\end{wrapfigure}
For outside access all nodes are virtualized to a degree at which the whole circle presents itself to the client as a single virtual server. When a client desires to perform an operation on the cloud he will - through respective routing mechanisms (load balancing) - be allocated to one specific node in the circle. Any \textit{valid} node can receive requests and depending on their current system load and geographical location the most suitable node for the client's request will be selected. A client can request \textit{read} operations and \textit{write} operations. These requests will be sent to the node and will either be confirmed or rejected within $\Delta$.  The internal communication between nodes consists of \textit{read} operations and \textit{write} operations, for which different protocols apply, and additional maintenance messages. The node a client is connected to will later be referred to as the \textit{triggering node}, as it is in charge of an operation and initiates the operation in the circle. The remaining nodes are referred to as \textit{not-triggering nodes}.

\subsubsection{Read:}
A \textit{read} consists of two types of messages. A \textit{request} message and a \textit{hash} message. The \textit{request} is a \textit{one-to-many} message from the \textit{triggering node}. The \textit{hash} message is the respective reply and is send by every \textit{not-trigering node} to the \textit{triggering node} as a \textit{one-to-one message}.

\subsubsection{Write:}
\textit{Writes} are performed based on the \textit{Two-Phase Commit Protocol (2PC)}\cite{2pc}. They consist of a \textit{write request} message, a \textit{confirmation} message and a \textit{commitment} message (Figure 6). Write requests and commitment are sent by the triggering node as a \textit{one-to-many message}, the confirmation messages are sent by the not-triggering nodes as \textit{one-to-one} messages.

\subsection{Algorithm}
\subsubsection{Write.}
A client connected to one of the nodes in the circle can perform \textit{write} operations. For this purpose the client sends a message containing the object he is wishing to write to the node. The message is not immediately confirmed to the client. The objects will first need to be successfully written on a majority of all nodes in the circle before the triggering node will confirm the operation to the client. This kind of majority mechanism is called quorum\cite{quorum}.
\begin{wrapfigure}{r}{0.5\textwidth}
  \vspace{-30pt}
  \begin{center}
\includegraphics[width=65mm]{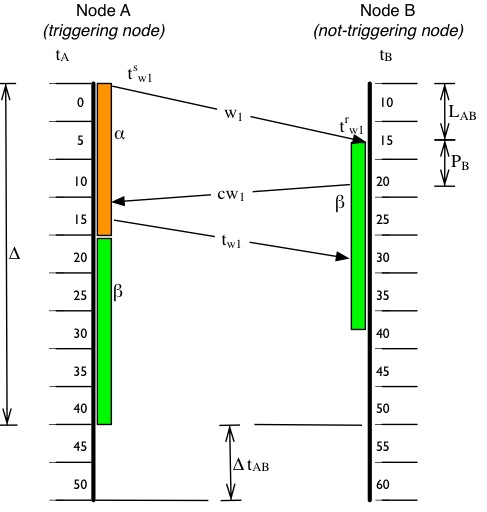}
  \end{center}
  \vspace{-20pt}
\caption{Event $w_1$ observed on nodes \textit{A} and \textit{B} $cw_1$}
  \vspace{-10pt}
\end{wrapfigure}

Let's first take a look at a simplifying example of two nodes to describe the basics of the interaction between triggering node and not-triggering node (Figure 7).

In every distributed system clocks are never perfectly synchronized and differ in speed\cite{TB}. The degree of retardation of the two clocks in the nodes \textit{A} and \textit{B} caused by their tendency of drifting apart is expressed by  \textit{$\Delta t_{AB}$}.  Though \textit{$\Delta t_{AB}$} exists from the point of view of an outside observer it can not be exactly measured. The uncertainty concerning the transit time of messages prohibits any exact measurement.
When messages are exchanged between the two nodes \textit{A} and \textit{B} there are two relevant timestamps. As shown in the example of \textit{Figure 7} a message \textit{$w_1$} leaves \textit{A} at time \textit{$t^s_{w_1}$}=0 and arrives at node \textit{B} at $t^r_{w_1}$=17. The letters \textit{s} and \textit{r} standing for \textit{send} and \textit{receive} respectively. The latency of $w_1$ being send from \textit{A} to \textit{B} is called $L_{AB}$ and can be calculated as:

$$L_{AB} = t^r_{w1} - t^s_{w1} - \Delta t_{AB}$$

By the definition of TSC every write on a node needs to be in the same order on every other node after $\Delta$ units of time. The algorithm handles this requirement in the following way:
Triggering node \textit{A} sends a message \textit{$w_i$} to all other nodes. The message \textit{$w_i$} contains the object of the write. A not-triggering node \textit{B} receiving \textit{$w_i$} will store the object with a temporary timestamp of $t^r_{w_i}$, which corresponds to the time the message \textit{$w_i$} arrives at the node \textit{B}, expressed in the local time of node \textit{B}. Node \textit{B} than confirms writing \textit{$w_i$} by sending back a confirmation message \textit{$cw_i$} carrying the timestamp $t^r_{w_i}$.
After receiving confirmations \textit{$cw_i$} of all four \textit{not-triggering nodes}, but at the latest $\alpha$ after sending out the initial write request \textit{$w_i$}, node \textit{A} commits to a timestamp it computes from the different \textit{$cw_i$} replies (See below: \textit{Commitment}). This committed timestamp is send as a message $t_{w_i}$ to all other nodes. The \textit{not-triggering nodes} receive the committed timestamp $t_{w_i}$ and overwrite the stored object's temporary timestamp with it. The data object is thereby becoming finalized.

To ensure that all requirements of TSC are met, certain conditions need to be fulfilled by the nodes to make the write successful. TSC requires that the system is either consistent after $\Delta$ units of time or becomes unavailable. The mechanism to handle these requirements relies on time windows. The \textit{triggering node} has two time windows of length $\alpha$ and $\beta$ respectively. $\alpha$ and $\beta$ sum up to exactly the maximum allowed treshold of time $\Delta$:

\[
\displaystyle  \alpha + \beta = \Delta
\]

The \textit{triggering node's} first time window $\alpha$ starts the moment the client 's write request is received (\textit{Figure 7}). As soon as a \textit{not-triggering node} receives the write request \textit{$w_i$} it opens a time window of size $\beta$. 
After the \textit{triggering node} receives confirmations ${cw_i}$ of all nodes and sends the committed timestamp $t_{w_i}$, but at the latest after $\alpha$ it starts it's second time window which also has a length of $\beta$.

The time window $\alpha$ has the purpose of limiting the maximum permissible time for the messages to transfer to the \textit{not-triggering node}, to be processed there ($P_B$) and to be confirmed to the \textit{triggering node}. In an ideal case of exactly constant network latencies ($L_{uv}$) and constant processing time for the request on the nodes ($P_v$) the required length of $\alpha$ can be computed as:

\[
\displaystyle \alpha = 2 L_{uv} + P_v
\] 

Determining $\alpha$ in a more realistic environment of falling and unpredictable networks requires a much more complex solution customized to the cloud environment and specific requirements of the client. Such a solution would need to take into account the items 1-6 of the list depicted in \textit{3.3 Algorithms}. 

It is in any case not necessary to forecast $L_{uv}$ or $P_v$ to guarantee TSC. Consistency properties are always guaranteed. It just improves performance if $\alpha$ is set to an adequate value. (See \textit{Choosing Parameters $\alpha$ and $\beta$})

Time window $\beta$ has the purpose of ensuring that a \textit{not-triggering node} is either consistent $\Delta$ after the initial triggering of a write or gets \textit{invalidated}. Nodes that are invalid do not react to any other reads or writes until they are valid again. When a \textit{not-triggering node} receives a write request $w_i$ the time window $\beta$ opens. The object is written with a temporary timestamp of $t^r_{w_i}$. It is now required that the committed timestamp $t_{w_i}$ is received within this time window $\beta$ to replace the object's temporary timestamp with the committed timestamp.
If the committed timestamp $t_{w_i}$ is not received within $\beta$ the node invalidated itself.

The instant of time a write confirmation \textit{$cw_i$} is send by a \textit{not-triggering node} is called $t^s_{cw_i}$, the moment the \textit{triggering node} receives it is called $t^r_{cw_i}$. From the perspective of an external observer it can obviously be concluded that $t^s_{cw_i}$ happens earlier than $t^r_{cw_i}$. 

Now we assume $cw_i$ is received by the \textit{triggering node} before it has committed to a timestamp and conclusively is still in time window $\alpha$. Due to the fact that $cw_i$ has been send by the \textit{not-triggering node} when it was already in its time window $\beta$ it can be concluded that the time window $\beta$ of the \textit{not-triggering node} starts \textit{before} the time window $\beta$ of the \textit{triggering node} does. Now that we know that the \textit{not-triggering node} either receives a committed timestamp within its time window $\beta$ or becomes invalid the final conclusion can be reached. A \textit{not-triggering node} receiving the write request $w_i$ will always either be consistent with it before $\Delta$ has passed since the client's initial request or become invalid. TSC is archived for all these nodes. Let's look at the possible, undesired outcomes in a realistic system that does not provide constant and reliable connections:

\paragraph{Case 1 - Unreachable nodes.}
A node can become unreachable for the write requests of the \textit{triggering node} for multiple reasons like network partition, invalidation or a system crash. In these cases the node will not be able to process writes or reply to messages and thus be inconsistent. However, the system as a whole can be consistent without requiring all of its nodes to be consistent. This is ensured by the quorum requiring more than 50\% of all nodes approving a write. To make sure the nodes don't stay inconsistent permanently the nodes \textit{heal}. The details of this mechanism will later be shown in detail.

\paragraph{Case 2 - $cw_i$ Arriving During $\beta$.}
A case where the relation of the \textit{not-triggering node's} $\beta$ always being previous to the \textit{triggering node's} $\beta$ is not guaranteed is when $cw_i$ arrives at the \textit{triggering node} later than $\alpha$. This case is possible but does not cause any complications because the \textit{triggering node} ignores any confirmations after $\alpha$. The committed timestamp will already have  been distributed to the timely responsive \textit{not-triggering nodes} at the end of time window $\alpha$. The \textit{not-triggering node} that has send the late confirmation message will conclusively not receive any committed timestamp and thus become unavailable.

\paragraph{Case 3 - $L_{uv}$ Being Longer Than $\alpha$.}
It is possible that the latency $L_{uv}$ of a write message $w_i$ is \textit{larger} than time window $\alpha$. This would also result in breaking the guarantee of one $\beta$ being before the other and as a result break the TSC guarantee. In the section \textit{Convergence} it will be shown how the clocks of the different nodes synchronize. Although perfect synchronization is not possible certain bounds can be defined. In his 1989 paper Flaviu Christian \cite{christian} shows that it is possible to synchronize clocks to any accuracy $\gamma$ (maximum permitted deviation of the two observed clocks) larger than half the round-trip-time (time required for a message to be sent from one point to another and back). 
If now a \textit{not-triggering node} receives the message $w_i$ at $t^r_{w_i}$ and the accuracy $\gamma$ is guaranteed in the system the latency of a message can me narrowed down. The latency $L_{uv}$ of a message received at $t^r_{w_i}$ can be isolated to the following interval:

 $$L_{uv}\in[\max\{0, t^s_{w_i} - t^r_{w_i} + \gamma \} , \max\{0, t^s_{w_i} - t^r_{w_i} - \gamma \}]$$
 
 To keep the sequential order of the time windows $\beta$ on the \textit{triggering-} and \textit{not-triggering} node it is now required that: $${L_{uv}\stackrel{!}{\le}\alpha}$$
 
Any message for which the maximum possible value of $L_{uv}$ is smaller than $\alpha$ will be accepted for a write as the sequential order of the time windows $\beta$ is guaranteed. Any message for which the maximum possible value exceeds $\alpha$ is rejected and the object will be locally invalidated.
The formal definition for a valid write is:

 $$ \alpha\stackrel{!}{\ge}\max\{0, t^s_{w_i} - t^r_{w_i} + \gamma, t^s_{w_i} - t^r_{w_i} -\gamma\}$$

Hence it is necessary that $\alpha$ is chosen sufficiently large in consideration of the exogene performance indicator $\gamma$.

\paragraph{Choosing Parameters $\alpha$ and $\beta$.}

The requirement initially defined was: 

$$\Delta=\alpha+\beta$$

When choosing values for $\alpha$ and $\beta$ it is highly recommendable to choose them such that:

$$\alpha\ll\beta$$

In the opposite case any write request $w_i$ having lower latency than the committed timestamp $t_{w_i}$ would result in exceeding the \textit{not-triggering node's} time window $\beta$ and thus in an invalidation of the node.

The optimal value of $\alpha$ and $\beta$ needs to be customized individually for every system to maximize performance. They will need to be defined depending on the variance of the network's latencies and customer's specific system requirements.

\paragraph{Commitment.}

TSC requires that after $\Delta$ an operation is sequentially consistent on all nodes. This means all events need to be in the exact identical order, defined by their respective timestamps, on all nodes. To be able to give this guarantee the algorithm needs to agree on a point in time it wants to place a write operation. As long as $\beta$ has not passed the write operation is not finalized on the respective nodes and the write's timestamp can still be changed during $\beta$. As soon as the committed timestamp $t_{w_i}$ is received by a \textit{not-triggering node}, the write can be finalized with the date of the committed timestamp.
Some nodes will receive the write $w_i$ at an earlier time $t^r_{w_i}$ than others. If now the committed timestamp $t_{wi}$ would be lower than the timestamp $t^r_{w_i}$ of a certain node, it would mean that this certain node would have to place the write operation far into the past, at an earlier point in time than the beginning of it's own time window $\beta$. As this would corrupt the maximum threshold $\Delta$ of TSC, the committed timestamp needs to be higher than the maximum of the timestamps $t^s_{w_i}$ and $t^r_{w_i}$ of the nodes:

\[
\displaystyle\textit{$t_{w_i} > \max\{{t^s_{w_i}, t^r_{w_i}} \}$}
\]

This way a node will never have to store a write operation dating back in time. Unfortunately it is now possible that the committed timestamp is one that does not fall into the time window $\beta$ on all nodes, but further in the future. As this problem can not be avoided without manipulating the clocks on respective nodes and thus corrupting any parallel process, the invalidation of certain nodes needs to be considered. To avoid a high number of nodes being invalidated $t_{w_i}$ needs to be chosen such that it respects the equation above and a maximum number of nodes stays valid. A respective validity-maximizing algorithm needs to be applied.

\paragraph{Quorum.}
For the write to be successful it is required that at least $\frac{3}{5}$ (or $\frac{n+1}{2n+1}$) of all nodes confirm the operation within the time $\alpha$ to the triggering node of the write. It is important for a majority of nodes to be consistent to avoid that for a later read a majority supports a "wrong" version. If at least 2 (or $n$) not-triggering nodes reply to the a message \textit{$cw_i$} to the triggering-node within the time window $\alpha$ the write is marked as successful and the client is informed. The committed timestamp $t_{w_i}$ is now send to all responsive nodes.
\begin{figure}[p]
\includegraphics[width=190mm, angle=90]{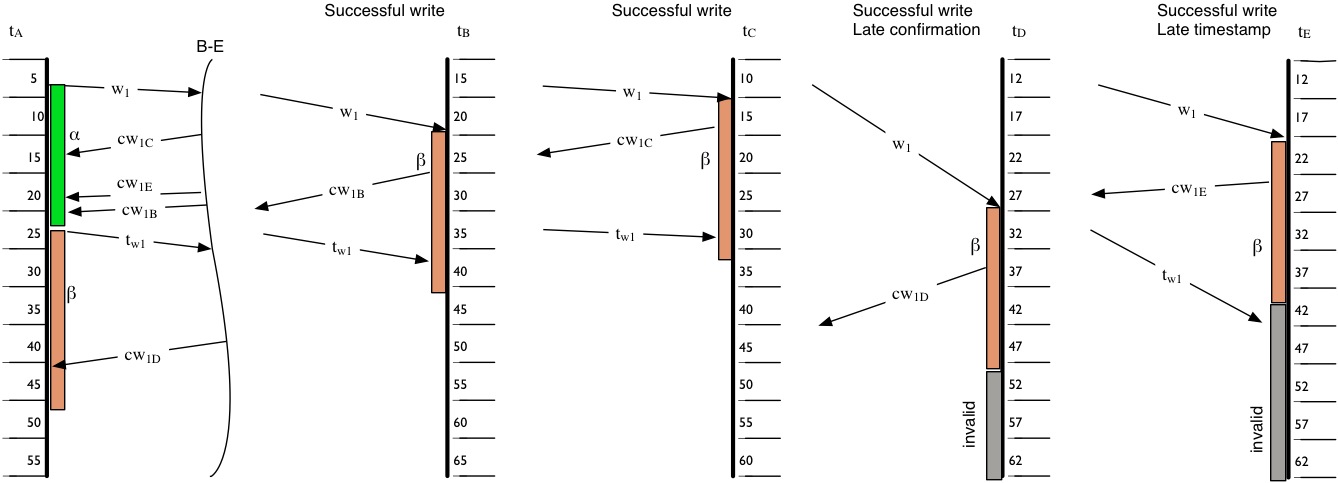}
\caption{Event $w_1$ observed on nodes \textit{A-E} with different failure conditions}
\end{figure}

\paragraph{Parallel Operations.}
A system will be required to handle multiple independent \textit{write} operations simultaneously. To ensure that TSC is also met in an environment of parallel operations, dependencies between the operations need to be avoided. To ensure this the \textit{write} operations are capsuled. They do not directly interact with one another. They share certain resources which have an influence on the algorithm. These resources therefore require strict control. The clocks are one of those resources multiple processes share. A \textit{write} operation is never allowed to manipulate the clock in any way. 
The processing speed on the servers and the speed messages are transferred on the network also affects the way the algorithm behaves and is a way multiple processes can indirectly influence one another. To avoid any such influence the algorithm takes processing time and network latency as exogenous. This effectively capsules single operations and effectively handles variations on the server and network infrastructure. Therefore any operation on the system can be considered isolated and independent.

\paragraph{Conclusions on Consistency.}
It can be concluded from the above that the algorithm implements TSC:

\begin{definition} 
\item The triggering node opens time window $\alpha$ after it writes \textit{$w_i$} .
\end{definition}
\begin{definition} 
\item All not-triggering nodes open their time window $\beta$ the moment they receive \textit{$w_i$}.
\end{definition}
\begin{definition} 
\item Only confirmation messages $cw_i$ that are received within $\alpha$ are being responded to with a  committed timestamp.
\end{definition}
\begin{definition} 
\item All messages have non-negative latencies.
\end{definition}

\begin{corollary}
\item The time window $\beta$ at any not-triggering node starts \textit{after} the time window $\alpha$ at the triggering node has started.
\end{corollary}

\begin{corollary}
\item The moment the triggering-node's time window $\beta$ ends all responsive not-triggering nodes will have passed their time window $\beta$.
\end{corollary}

\begin{corollary}
\item Conclusively all nodes that received \textit{$w_i$} either received \textit{$t_{w_i}$} and are sequentially consistent with respect to \textit{$w_i$} or are invalidated before the triggering nodes's time window $\beta$ ends.
\end{corollary}

\subsubsection{Read.}
When a client is connected to a node \textit{A} of the circle and requests to read an object it is required that at least $\frac{3}{5}$ (or $\frac{n+1}{2n+1}$) of the nodes agree on the content of that piece of data. This ensures that at least 50\% of all nodes agree and tolerates up to $\frac{2}{5}$ (or $\frac{n}{2n+1}$) of the nodes being unavailable. What the triggering node will have to do in this case is identify two other nodes that store the exact same copy of the requested object the triggering node itself is storing. Therefore node \textit{A} sends a message to all the other nodes requesting them to return the \textit{hash value} of the requested object. As soon as 2 (or $n$) nodes have replied a hash value identical to the one \textit{A} has generated from it's locally stored object, \textit{A} sends the requested data object to the client. 

As the implementation of TSC is supposed to offer an end-to-end consistency guarantee it appears intuitive to also give a guarantee on the time necessary to complete a \textit{read} operation. We have seen that it is possible to define a $\Delta$ for which data-centric consistency is established, corresponding to the $\Delta_{Data}$ referred to early in section 3. Now that we say that client-centric consistency is only established if timed consistency is experienced from a client point of view the $\Delta_{Client}$ and $\Delta_{Network}$ still need to be determined:
$$\Delta_{Update}=\Delta_{Data} + \Delta_{Client} + \Delta_{Network}$$

But as seen in Figure 4 the propagation time of an update form one client to another is not limited only by the transaction speed and access time. It also depends on the point in time the object is being accessed. Any update by one client \textit{can} be received by any other client after $\Delta_{Client}$, which does not necessarily mean it will immediately be. If the client performs \textit{if-modified-since} request in regular intervals as proposed by Torres-Rojas et al. the $\Delta_{Client}$ is simple equal to the frequency of requests. 
Now that only $\Delta_{Network}$ is left to be determined it is becoming more obvious that the distributed system will not need to guarantee an upper bound for the system to return a read request. $\Delta_{Network}$ is a time that can only be measured from the clients point of view as only he is aware of the exact time between sending the read request and receiving the answer. 
Now that the distributed system guarantees a certain time window $\Delta_{Data}$ and the client is aware of it's request intervals $\Delta_{Client}$ the guarantee is complete.

If $\Delta_{Network}$ is \textit{within} the maximum allowed delay the client defined, an object is considered consistent. If $\Delta_{Network}$ is exceeded the $\Delta_{Update}=\Delta_{Data} + \Delta_{Client} + \Delta_{Network}$ condition is broken and the read data is being considered stale.

As we have just seen the data-side gives guarantees on the timeliness of the data provided, but the final information whether read data is stale or not can only be made from the client's point of view as the last link in the chain. 
Now that there is no mandatory upper bound required by TSC for the time the data-side needs to respond to a \textit{read} request, it can be evaluated whether it is reasonable to define one for other reasons.

Though it is not required to guarantee TSC it might very well make sense to define an upper bound of time for read requests, too. The client might be interested to know about the condition of the data-side. If the triggering node is suffering from network partition and is allowed to take indefinite time to respond to read requests, the partition will never come to the client's attention. The same argument is valid for the operator of the distributed system who will need an upper bound for the time a read needs to execute to detect problems within the system and to proove compliance with the SLAs.

The upper threshold of time for a read request to be either returned or rejected will be defined by  $\omega$. After $\omega$ a read request from a client either returns a value or gets rejected.

~\\
There are the following possibilities for abnormalities in the algorithm that have to be considered in detail:

\begin{enumerate}
\item{At the threshold time of $\omega$ after the triggering node requested the hash code, less than 2 nodes have replied an identical hash code to the one \textit{A} generated and less than 3 nodes have send identical hash codes to one another. In this case the read fails and \textit{A} replies to the client that the read was not successful.}
\item{If at $\omega$ 3 not-triggering nodes \textit{did} reply identical hash codes, but these do not match to the one the triggering node generated, it will return the object to this majority-compatible hash as a response to the client and invalidate itself to heal it's own inconsistency.}
\item{At $\omega$ at least two nodes have send hash codes identical to the one \textit{A} has generated but 1 or more nodes have replied different hash codes. \textit{A} will for all nodes replying invalid values, mandate the next neighboring node counter clock-wise with the correct reply to heal this inconsistency. The mechanism is shown below in section \textit{Anti-Entropy}.} The successful delivery of this request from \textit{A} to the respective node is desirable but not required to secure the condition of TSC. This explains why no confirmation is required. Even if the message is lost, consistency is secured by the quorum which ensures that at least 3 nodes always carry the correct data and that even if all of them fail the remaining two can not overrule them. The reason for the healing to take place nevertheless is availability. A node carrying stale data increases the chance of a read that can not be complete because the quorum does not find 3 identical copies in time.
\end{enumerate}

\subsubsection{Anti-Entropy.}
A triggering node can request another node to heal the inconsistency of it's clockwise neighbor. In this case the node receiving such a request will start an \textit{Anti-Entropy}\cite{anti-entropy} protocol to synchronize all replicas. To reduce the amount of data being exchanged by comparing each item's hash code consecutively the algorithm uses a \textit{Merkle Trees}\cite{merkle} (See also \textit{Apache Cassandra}\cite{cassandra} and \textit{Amazon Dynamo}\cite{ADynamo} ). A \textit{Merkle Tree} is a hash tree where the leaves correspond to objects. Generally a \textit{Merkle Tree} is a binary tree, meaning two leaves are children to one father node. Father nodes are hash values generated from its children's values. The advantage of a \textit{Merkle Tree} is that there is no need to check every leave. Instead data structures can be checked by comparing the root hash key and moving down the tiers till the inconsistent leave is successfully identified.
After the root hash key of the \textit{Merkle Tree} is consistent between the two involved nodes the \textit{Anti-Entropy} is finished successfully.

\subsubsection{Convergence.}

Although Anti-Entropy heals inconsistent nodes once they are detected it is possible for nodes to be inconsistent and not be detected. This is the case in a system which stores data that is hardly read at all. Take for example a circle which is used just for storing replicas for means of redundancy. The data will in the average case not be accessed at all. If one day the data \textit{is} required it is possible that the data on the hard drives changes due to physical damages. Now if the data is not accessed for a very long time and thus hasn't been compared for quorum means it is possible, although not very probable, that data on more than 2 nodes is damaged and the circle itself is thereby permanently invalid. To avoid this, while not reducing performance of a system, a node can use times of longer inactivity in the circle to trigger a preventive \textit{Anti-Entropy} that moves around the circle. This efficiently ensures Convergence. A clockwise example is depicted in \textit{Figure 9}. 

\begin{wrapfigure}{r}{0.5\textwidth}
  \vspace{-30pt}
  \begin{center}
\includegraphics[width=70mm]{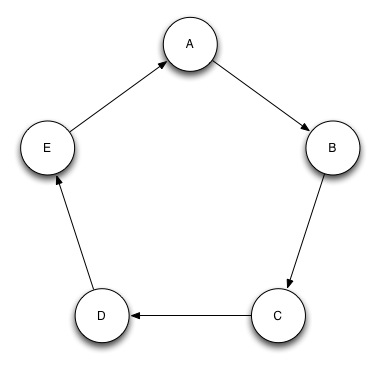}
  \end{center}
  \vspace{-20pt}
\caption{Convergence}
  \vspace{-10pt}
\end{wrapfigure}

\subsubsection{Revalidation.}
Nodes invalidate themselves for four reasons. Either they have processed a write but did not get the finalizing timestamp before the time window $\beta$ has passed (see Section \textit{Write}), they have generated an inconsistent result in a read, they have detected an internal server error causing the system to hold or they have just recently been added to the circle and are not consistent yet. In either case the system will try to get consistent and valid again. To archive this a node can simply request the \textit{Anti-Entropy} protocol by his counter clockwise (ccw) neighbor. As soon as the \textit{Anti-Entropy} is completed a node will set its internal status to \textit{valid} and begin to handle client request, as well as requests by other nodes in the circle. As it is possible that the ccw neighbor is invalid itself it will repeat this request continuously till \textit{Anti-Entropy} is executed. A chain of nodes being unavailable will therefore become available one by one in a ccw order.

\subsubsection{Clock Synchronisation.}

As shown earlier the clocks of servers and consequently the nodes on them are never perfectly synchronous and they drift apart due to their difference in speed. In the case of this implementation it is acceptable to neglect the actual difference in the clock's speed \textit{during} a certain observed action. If two nodes are being observed for the time $\Delta$ the clocks will drift apart slightly but as $\Delta$ itself is an extremely short time window this effect can be neglected. However the difference in speed causes the clocks to drift apart in the \textit{long run}. This effect has to be considered as it can affect the asynchrony of clocks substantially. 
For the implementation of any flavor of sequential consistency it is necessary that events can be given an order. This order needs not reflect the actual realtime order as an outstanding observer would see it. It is only important that the nodes agree on the order of events. 

As seen in \textit{"Write - Case 3"} clocks can be synchronized to any maximum deviation $\gamma$, which needs to be higher than half the round-trip-time\cite{christian}. To make sure that the deviation of two clocks $L_{uv}$ stays below $\gamma$, synchronization has to take place in regular intervals. The exact frequency of those synchronizations depends of the speed of the continuous apart drifting of the clocks.

\subsection{Results}

The previously described implementation produces guarantees for the customer that are summed up in the following list:

\begin{enumerate}
\item{$\Delta$ units of time after one client performs a \textit{write}, this write will be in the exact sequential order compared to other writes for any client accessing it \underline{or} the write is being \textit{rejected} after $\alpha$.}

\item{Therefore\textit{Timed Sequential Consistency} can be guaranteed.}

\item{At the latest $\omega$ units of time after a client performs a \textit{read} it will return a consistent result \underline{or} be rejected.   }

\item{Permanent \textit{Session Consistency} (see \cite{Vogels}) can be guaranteed.}

\item{Permanent \textit{Monotonic Read Consistency} (see \cite{Vogels}) can be guaranteed.}

\item{Permanent \textit{Monotonic Write Consistency} (see \cite{Vogels}) can be guaranteed.}

\end{enumerate}

\section{Feasability}

The feasibility of the suggested solution will be assessed by a number of performance indicators that have an influence on the customer. Whether or not a customer will apply TSC depends on the match between those performance indicators and the customer's needs:

\subsubsection{Reliability}
\begin{enumerate}
\item{Availability}
\item{Correctness of Reads}
\item{Loss of Data}
\end{enumerate}
~\\
\textit{Availability: }Depending on the length of $\Delta$ (more specifically $\alpha$ and $\beta$)  and $\omega$ the success rate for writes and reads varies and thereby \textit{availability} is determined. The variables can be configured to match the desired availability properties. The exogenous factors of availability can be influenced by the performance of the data center hardware, such as the latency of servers and network components.

~\\
\textit{Correctness of Reads: } The guarantee is \textit{very} close to 100\%. $\frac{3}{5}$ ($\frac{n+1}{2n+1}$) would need to return data that mutated exactly in the same way to breach the correctness criteria which is $extremely$ improbable.

~\\ 
\textit{Loss of Data:} It is most unlikely that data is completely lost, because redundancy of data is always higher or equal to $\frac{3}{5}$ $(\frac{n+1}{2n+1})$.

\subsubsection{Efficiency}

\begin{enumerate}
\item{Network Load}
\item{System Load}
\item{Data Volume}
\item{Cost}
\end{enumerate}
\textit{Network Load: }Under failure free conditions the network load in the circle can be expressed as:

Write: 6n = 2n [Object] + 4n [Timestamps]

Read: 4n = 2n [Object ID] + 2n [Hash]

~\\
with $n$ being the factor for the number of nodes (e.g. 2n+1).

~\\
\textit{System Load:} The load generated on the servers is relevant and can be subject to further research in this topic. It can not be assessed more deeply in this paper because specific software architecture is not in the scope of this paper.

~\\
\textit{Data Volume: }The total amount of data stored in the distributed system is comparably high due to the high degree of replication \textit{N=5} in the recommended configuration, in contrary to a replication of \textit{N=3} generally used at Amazon data centers\cite{Vogels}.

~\\
\textit{Cost:} For the great majority, the cost of the infrastructure depends on the required hardware and energy consumption, and thus in the end dependents on \textit{network load, system load and data volume} (for further details on cost structures see \cite{BER}).

\subsubsection{Maintainability}

\begin{enumerate}
\item{Failure Detection}
\item{Regular Maintenance}
\item{Replacement}
\item{Healing}
\end{enumerate}

~\\
\textit{Failure Detection:} The detection of failures is $not$ an integral part of the algorithm because the quorum can not differentiate between a wrong value due to \textit{inconsistency} and one due to \textit{damaged data}. Neither will the node itself be aware of any wrong value it delivered because there is no feedback to the node. On the other hand the system itself takes care of \textit{non-permanent} failures just as it does with inconsistencies. \textit{Anti-Entropy} and \textit{Convergence} can efficiently detect such failures and remove them, without the danger of a failure spreading (see Quorum).

~\\
\textit{Regular Maintenance: }Removing nodes from the circle to perform maintenance is not critical to the operation as this corresponds to network partition which can be handled by a system with appropriate $\Delta$. 

~\\
\textit{Replacement:} Physically replacing the hardware of a node can also be handled as it corresponds to a network partition with a subsequent massive inconsistency which will be handled by \textit{Anti-Entropy} and \textit{Convergence}. 

~\\
\textit{Healing:} For many kinds of scenarios, as described in earlier sections, healing is taken care of by \textit{Anti-Entropy}, \textit{Convergence} and \textit{Clock Synchronization}.

\subsubsection{Latency}

\begin{enumerate}
\item{Network}
\item{System}
\item{Data Storage}
 ~\newline{}
 
Write: 
\begin{enumerate}
\item{\textit{Successful write:} 	

$$L_{Write}=\max{\{L^w_3\}}$$

with $L^w_3$ being a set of the three shortest latencies $l^w_v$ for a write request and confirmation being transferred to the respective node $v$ and back.

~\newline
 $l^w_v=l_{w_v}+l_{cw_v}+P_B$} , $v\in N$ 
 
 ~\\
 $l_{w_v}$: Latency for write request being send from triggering node to node $v$.\\
 
 $l_{cw_v}$: Latency for a confirmation message being send from node $v$ to the triggering node.
 ~\newline

\item{\textit{Sequentially consistent write:} 
$$L_{Write}=\alpha+   \max\{{L^{c}_3,}\}<\Delta $$

with $L^c_3$ being a set of the three shortest latencies $l^c_v$ for a committed timestamp $t_{w_v}$ being transferred to the respective node $v$.

 }
\end{enumerate}

 ~\newline{}

Read:

$$L_{Read}= \max{\{L^r_3\}}<\omega$$

with $L^r_3$ being a set of the three fastest response times $l^r_v$ for a hash request to node $v$.

~\\
$l^r_v=l_{hr_v}+l_{h_v}$ , $v\in N$ 

~\\
$l_{hr_v}$: Latency for a \textit{hash request} message being send to node $v$.\\

$l_{h_v}$: Latency for a \textit{hash} message being returned by node $v$.

\end{enumerate}
~\\
For well configured $\alpha$ and $\beta$ and hardly variant latencies, TSC can reach comparable \textit{write latency} as a system configured for Weak Consistency with (5,3,3) setup. Weak Consistency systems with lower $N$ (generally 3, see \cite{Vogels}) will have lower write latencies, assuming equal exogenous factors.
 
 ~\\
\textit{Read latency} will behave exactly equal to a system configured for Weak Consistency and same (N,R,W) setup. Setup (3,2,2), more frequently used with Weak Consistency\cite{ADynamo}, will outperform any TSC configured system of (2n+1,n+1,n+1).

\subsubsection{Scaleability}

\begin{enumerate}
\item{Network Load}
\item{System Load}
\item{Stored Data}
\item{Cost}
\end{enumerate}
~\\
\textit{Network Load:} The internal \textit{Network Load} generated by writes and reads behaves linear to the number of nodes. The number of internal messages produced by \textit{writes} is $6n$ and for \textit{reads} 4n respectively.

The first reaction to high write/read demand is \textit{Horizontal Scaling}, meaning that the number of \textit{different} pieces of data in one circle is reduced by moving some data to a new circle. This reduces the pieces of data a circle has to handle and thus reduces loads. TSC will hold between the items non the less. This scaling mechanism has an upper bound at the point where a circle only handles a single piece of data. 

Beyond this point scaling is possible by changing the physical \textit{node allocation}. The number of nodes hosted on one physical machine (server) can be reduced to the point where only one node is hosted on a physical machine. \textit{Horizontal Scaling} combined with changed \textit{node allocation} can take the system to a point where every node handles a $single$ piece of data on $one$ dedicated server.

For the very view applications that are insufficiently served by this solution further scaling is only possible by improving exogenous factors, such as hardware performance.

~\\
\textit{System Load:} The performance requirements on the system can be expected to be linear to the number of write/read requests $i$. For a vertically scaled system (increase in $N$) the system load can be expected to grow below linear due to synergy effects in the messages sent and received.

~\\
\textit{Stored Data:} The amount of unique customer data to be stored, that guarantees TSC, can be scaled horizontally with $linear$ effort in hardware. Twice the amount of data will require twice as many nodes and thus twice as many servers at constant availability.

~\\
\textit{Vertical Scaling:} Increasing the number of nodes $N$ produces a linear increase in the total amount of data stored, as the number of replicas increases.

~\\
\textit{Cost:} As the costs for the cloud infrastructure are for great parts hardware and energy costs, they can be expected to be slightly below linear to the amount of servers, as economies of scale apply. For more details see \cite{BER}.

\subsubsection{Adaptability}

\begin{enumerate}
\item{Duplicates}
\item{Latency}
\item{Application Types}
\item{Desired Upper Bound $\Delta$}
\end{enumerate}

~\\
\textit{Duplicates:} The number \textit{N} of duplicates can be changed (Vertical Scaling). This is possible during operation but requires certain time. The number of nodes can only be changed in even numbers. If such a mechanism is required, the necessary changes can be made in the \textit{Convergence} mechanism.

~\\
\textit{Latency:}  The latency of a distributed system can be adjusted by providing more resources for an operations. \textit{Horizontal Scaling} is such a mechanism. As shown in \textit{Scaleability} there is an upper bound after which only improving exogenous variables, such as network and server speed, will improve latency.

~\\
\textit{Application Types:} The suggested solution for TSC does not provide an adequate solution for a read-centric application, like the N=100, R=1, Vogels quotes in his article\cite{Vogels}. While $\frac{N}{R}$ is normally very high for read-centric applications (100 in the previous example)  the suggested algorithm can only offer $\frac{N}{R}=\frac{2n+1}{n+1}$ which will converge to $\frac{N}{R}=2$ with high \textit{n}. This is not unexpected as TSC is suppose to be a very strong consistency criteria and read-centric applications are normally rather sloppy with consistency.
Applications using Strong Consistency, such as Google Drive\cite{ctypes} or banking applications, can obviously not migrated to a TSC system.

~\\
\textit{Desired Upper Bound $\Delta$: } The value for $\Delta$ can be individually defined for any circle by changing $\alpha$ and $\beta$ accordingly. As a circle can consist of as few as a single object, every object can have an individual $\Delta$. Objects that are supposed to have a TSC relationship to one another are required to have equal values of $\Delta$.

Depending on the value of $\Delta$, the tradeoff between \textit{availability} and \textit{latency} can be made within the performance borders of the systems (Network/System Latency). For the needs of the respective application, the optimal value can be decided on. 

\subsubsection{Descriptiveness}

~\\~\\
The aim of TSC expressed in the beginning of this paper was to establish a reliable upper bound of time for an update to travel through the system. This upper bound was suppose to give the developer a certain guarantee on the behavior of the cloud. The performance of a solution for TSC will as a conclusion only be as good as the reliability of the guarantee. Only if the customer knows exactly how fast TSC is established and how probable the availability is afterwards, it generates an added value. Therefore \textit{Descriptiveness} is a key indicator. It is important that the properties of TSC can be expressed clearly in a service level agreement (SLA) and that the compliance with the SLA can be precisely $measured$.

The \textit{Descriptiveness} of the suggested TSC implementation is comprehensive, meaning that there is total transparency on when consistency is established. TSC can effectively not be broken without breaking the system. If consistency can not be established the system will be rendered unavailable till consistency is restored. Therefore a SLA will not express the level of consistency, but the probability of a successful operation without rendering the system unavailable. The parameters that will have to be measured and agreed upon in contracts are: The probability of a successful write, the probability of an available system after a write with respective $\Delta$ and the probability of a read being successfully completed after $\omega$. These parameters can be measured in a transparent way and therefore are suitable for SLA's.

\section{Related Work}
Traditionally Strong Consistency has been the dominant paradigm in the research of distributed systems (Haerder 1983\cite{ACID1}, Bernstein 1984 \cite{Bernstein}). It was supplemented by the early approaches of optimistic systems(Kisteler et al.\cite{Kistler}) with rather limited impact.

With the rise of the CAP-Theorem (Fox et al. 1997 \cite{antiacid}, Brewer 2000\cite{CAP}) the idea of weakening consistency in order to increase availability and tolerance to network partition has been increasingly researched. While Strong Consistency keeps a supremacy in  certain fields of applications, weak consistency is applied in some of today's leading technologies in distributed systems, like Amazon Dynamo (Vogels 2008\cite{Vogels}, DeCandia et al. 2007\cite{ADynamo}), Apache Cassandra (Lakshman et al. 2010\cite{cassandra}) or the Google File System(McKusick et al.\cite{GFS})

Consistency models mostly focus on ordering events. Torres-Rojas et al.\cite{TR} instead emphasize the timeliness of consistency. They considering time as an endogenous factor rather than as a result of the selected consistency model. Based on the arbitrary timeliness requirement they attach a lifetime to an object and develop a theory about the consistency properties on its basis. Given a distributed system would exist, that itself establishes timed consistency from a data-centric point of view, an end-to-end client-centric consistency is implied by Torres-Rojas et al. model\cite{Lifetime}\cite{CTaWCM}. 

Developing such a data-centric consistency system and archiving the effects the model of Torres-Rojas et al. anticipates is the objective of this paper. It focuses on the data-centric propagation mechanisms necessary to archive data-centric timed consistency and describes how to make it experienceable by the client side mechanisms of Torres-Rojas et al.. The result is a consistency model that allows an integrated end-to-end timed consistency for interaction of clients (client-centric). 

In opposite to other tunable consistency approaches, Torres-Rojas et al. approach to timed consistency is strictly limited to the client side and implicitly makes assumption on the constitution of the data side of the distributed system. The interval of updates is the central parameter. It decides the maximum accepted age of an object and therefore defines the staleness of an object.
Yu et al. \cite{conit} in contrast allow for application specific consistency levels based on a  more complex three factor metric. 
An approach that also considers the data-centric consistency in detail is Krishnamurthy et al. \cite{Tuneable}. Their approach is based on individually tuning there data-centric consistency requirements for replicas to meet the requirements, based on probabilistic reasoning. In contrast to Krishnamurthy et al. the data-centric consistency  of this paper's model is flexible by object but strict in terms of uniform execution strategies and limitations. Strict execution guarantees and informational certainty are favored to probabilistic methods of increasing performance.

\section{Conclusions}

Solutions for distributed storage generally offer just \textit{Eventual Consistency}. The theoretical implementation described in this paper shows that it is possible to archive comprehensive \textit{Timed Sequential Consistency}. TSC can be implemented in a way that makes it client-centric and thereby covers all the information chain from one client performing the update, through the cloud, to any other client receiving it.

In an environment of failing networks the suggested implementation of TSC is able to guarantee \textit{consistency} regardless of the degree of network partition or system failure. The rather strong consistency properties are traded off for the availability which is weaker and tolerates just less severe network partitions or failures, compared to models of \textit{Eventual Consistency}. Nevertheless, the margin of tolerance in which a system will stay $available$ even though failures occur, is still close to 50\% of all servers being unreachable in time. For a system that has very predictable latencies and low variances in load, the availability can reach a level very close to the availability level on a system with \textit{Eventual Consistency}. The lower tolerance is for great parts compensated by a higher number of replicas in the proposed standard configuration (N=5), compared to normal distributed storage solutions. Therefore it can be concluded that the implementation for TSC is a comparable costly solution but offers very strong consistency properties, that availability wise can almost compare to an \textit{Eventual Consistency} system.

The application range for the TSC implementation is rather broad due to the flexible definition for the upper threshold $\Delta$. Highly read-centric applications are in most cases less relying on consistency and will therefore be better served by a weaker consistency model with lower latencies. While applications requiring \textit{Strong Consistency}, like many sensible banking and security related applications, can obviously not be satisfied by a \textit{Weak Consistency}\cite{CTaWCM} model like TSC, there is a wide range of applications, neither requiring extremely high read-loads nor Strong Consistency, that requirements can be met\cite{ctypes}. These customers in between the extremes of \textit{Strong Consistency} and loose \textit{Eventual Consistency} are addressed by TSC. Enterprise customers having the need for a more predictable consistency than \textit{Eventual Consistency} can offer, might find TSC tempting. 

Variations in the system load occur in most distributed systems. The TSC implementation offers the possibility to efficiently scale the system, within certain limits. Units of data can be moved to additional circles to a degree where a node, lying on a single server, manages only a single piece of data. Such a technique will drastically improve performance but $has$ certain limits. Beyond this point performance can only be increased by exogenous factors such as faster networks and lower latencies in servers, which impose a strong technological barrier. Even in such a situation consistency can be guaranteed \textit{across} circles due to $\Delta$ which can be applied to every circle in the same way. 

The implementation of TSC strongly emphasizes the relationship between \textit{enterprise customer} and \textit{cloud provider}. 
TSC permits customizing the cloud's performance properties to the specific requirement of a business customer. \textit{Availability, consistency and latency} can be configured through endogenous variables. The performance properties can effectively be agreed upon as a level of conformity in \textit{Service Level Agreements}. The implementation allows  to measure the respective variables efficiently and to transparently determine the level of compliance with the SLA. High transparency and wide predictability for the customer are the result.

For further research into the topic it will be interesting to alter the model of \textit{Timed Sequential Consistency} such that \textit{Timed Causal Consistency} can be reached. TCC offers the potential to improve availability and to decrease latencies while weakening consistency in a way that will be most tolerable for the majority of customers. The presented TSC implementation is designed in a way that will facilitate building such TCC solution.

~\newpage



\bibliography{bib/bib}
\bibliographystyle{llncs/splncs}




\section*{Supplementary material}

# Flowchart: Knot Protocol State Diagram

**States and transitions (read sideways):**

- **Stand-by**

### Write Path (Triggering knot)
- User performs Write on knot → **Triggering knot - Write**
- Send write command to all knots → **Waiting for confirmation**
  - ≥2 knots reply before Δ/2 → **Successful write**
    - ≥3 stamps are within Δ/2 → **Stamp selection**
      - Availability maximizing stamp is selected → **Knot information**
        - All knots matching stamp will be notified of committed stamp → **End of process**
    - <3 stamps are within Δ/2 → **Reversal**
      - All knots receive empty timestamp and user gets error message → End of process
  - <2 knots reply before Δ/2 → **Unsuccessful write** → End of process

### Write Path (Non-triggering knot)
- Other knot performs write → **Non-triggering knot - Write**
- Timestamp send to triggering knot → **Timestamp generation** → **Waiting for confirmation**
  - Timestamp received within Δ/2 → **Finalization** (Data written successfully) → End of process
  - Timestamp not received within Δ/2 → **Invalid knot** → Request Anti-Entropy from cow neighbor

### Read Path (Triggering knot)
- User performs Read on knot → **Triggering knot - Read**
- Send all knots request for hash code → **Waiting for confirmation**
  - ≥2 knots reply in time → **Anti-Entropy** → Anti-Entropy request send to cow neighbors of unresponsive knots → **Hash comparison needed**
    - ≥3 knots same hash → **Positive quorum** → **Anti-Entropy**
      - Triggering knot has correct hash → **Triggering knot true** → Send file to user → End of process
      - Triggering knot has incorrect hash → **Triggering knot false** → Handover to true hash knot → **Invalid object** → Self-Invalidation → Enforce Anti-Entropy
    - <3 knots same hash → **Negative quorum** → Send Anti-Entropy request to all knots → **Enforce Convergence** → End of process
  - <2 knots reply in time → **Anti-Entropy** → Anti-Entropy request send to cow neighbors of unresponsive knots → **Unsuccessful read** → Inform user of unsuccessful write → End of process

### Read Path (Non-triggering knot)
- Other knot performs read → **Non-Triggering knot - read**
- Generate hash code for requested file → **Hash generated** → Send hash to triggering knot → End of process

### Anti-Entropy - Active
- Anti-Entropy requested for cow neighbor → **Anti-Entropy - Active**
- Send root hash of Merkle-Tree → **Wait for comparison**
  - Hash identical → End of process
  - Different hash → Move down Merkle-Tree → **Identify inconsistency**
    - All inconsistent objects identified → **Merkle-Tree finished** → send respective objects → End of process
  - No reply → End of process

### Anti-Entropy - Passive
- Cow neighbor enforces Anti-Entropy by sending root hash → **Anti-Entropy - Passive**
  - Identical root hash of Merkle-Tree → End of process
  - Different hash → **Identify inconsistency**
    - Move down Merkle-Tree
    - All inconsistent objects identified → **Merkle-Tree finished** → Send respective objects → End of process

- End of process → Stand-by



\end{document}